\documentclass[aps,twocolumn,prm,reprint,amsmath,amssymb,superscriptaddress]{revtex4-1}
\usepackage[utf8]{inputenc}
\usepackage{amsmath}
\usepackage{amsfonts}
\usepackage{amssymb}
\usepackage{graphicx}
\usepackage{bm}
\begin{document}

\title{Near-thermal limit gating in heavily-doped III-V semiconductor nanowires using polymer electrolytes}

\author{A.R.~Ullah}
\affiliation{School of Physics, University of New South Wales, Sydney NSW 2052, Australia}

\author{D.J.~Carrad}
\affiliation{School of Physics, University of New South Wales, Sydney NSW 2052, Australia}
\affiliation{Center for Quantum Devices and Station Q Copenhagen, Niels Bohr Institute, University of Copenhagen, Universitetsparken 5, DK-2100 Copenhagen, Denmark}

\author{P.~Krogstrup}
\affiliation{Center for Quantum Devices and Station Q Copenhagen, Niels Bohr Institute, University of Copenhagen, Universitetsparken 5, DK-2100 Copenhagen, Denmark}

\author{J.~Nyg{\aa}rd}
\affiliation{Center for Quantum Devices and Station Q Copenhagen, Niels Bohr Institute, University of Copenhagen, Universitetsparken 5, DK-2100 Copenhagen, Denmark}

\author{A.P.~Micolich}
\email{adam.micolich@nanoelectronics.physics.unsw.edu.au}
\affiliation{School of Physics, University of New South Wales, Sydney NSW 2052, Australia}

\begin{abstract}
Doping is a common route to reducing nanowire transistor on-resistance but has limits. High doping level gives significant loss in gate performance and ultimately complete gate failure. We show that electrolyte gating remains effective even when the Be doping in our GaAs nanowires is so high that traditional metal-oxide gates fail. In this regime we obtain a combination of sub-threshold swing and contact resistance that surpasses the best existing $p$-type nanowire MOSFETs. Our sub-threshold swing of $75$~mV/dec is within $25\%$ of the room-temperature thermal limit and comparable with $n$-InP and $n$-GaAs nanowire MOSFETs. Our results open a new path to extending the performance and application of nanowire transistors, and motivate further work on improved solid electrolytes for nanoscale device applications.
\end{abstract}

\date{\today}

\maketitle

\section{Introduction}
Complementary metal-oxide-semiconductor (CMOS) technology is central to modern integrated circuits. The technology combines $p$-type and $n$-type Metal-Oxide-Semiconductor Field-Effect Transistors (MOSFETs), exploiting the opposing charge carrier polarity and channel current versus gate voltage characteristics to achieve low power logic. Miniaturisation drove development of nanowire CMOS featuring III-V nanowires integrated on Si towards high performance at low cost~\cite{WernerssonProcIEEE10, delAlamoNat11, RielMRSBull14}. This is underpinned by broader research aimed at improved gating in III-V Nanowire Field-Effect Transistors (NWFETs) seeking steeper sub-threshold slope and lower parasitic resistance for reduced operating voltage and enhanced energy efficiency~\cite{AppenzellerTED08}. Progress has been strong for $n$-type III-V NWFETs with near-thermal limit sub-threshold slope obtained for InP~\cite{StormNL11}, InGaAs~\cite{TomiokaNat12} and AlGaAs/GaAs~\cite{MorkotterNL15}. Integration of $n$-type III-V NWFETs on Si is well established~\cite{TomiokaNat12, SchmidAPL15} with GHz operation demonstrated~\cite{JohanssonEDL14}. Most of these devices achieve their excellent gate channel coupling by employing high-$\kappa$ dielectrics such as HfO$_2$.

The development of $p$-type III-V NWFETs has lagged behind~\cite{WernerssonProcIEEE10}. This is caused by several key challenges for $p$-type devices including lower intrinsic carrier mobility and difficulties in growth, doping and fabrication of high quality ohmic contacts and gates. Hence III-V nanowire CMOS typically features $p$-type transistors far less ideal than their $n$-type counterparts~\cite{StormNL11, DeyNL12, SvenssonNL15}. Here we present polymer electrolyte gated Be-doped $p^{+}$-GaAs NWFETs with near-thermal limit gating that point out a path to filling this significant performance gap. Our $p^{+}$-doping ensures low contact resistance and high channel conductivity while the electrolyte gate provides sub-threshold slope $\sim75$~mV/dec, on-off ratio of order $10^4$ and low hysteresis. This is despite the doping level being so high that traditional metal-oxide gate structures fail completely, i.e., give no gate modulation or very poor on-off ratio ($<3$)~\cite{UllahNanotech17}. In particular, even top gates with HfO$_2$ dielectric -- similar to those that facilitate near thermal limit switching in, e.g., InP~\cite{StormNL11} -- do not provide effective transistor action.

Our polymer electrolyte (PE) gate consists of an electron-beam patterned polymer gel, e.g., polyethylene oxide, spanning the gap between a metal electrode and the nanowire~\cite{CarradNL14}. The gel absorbs water providing an environment for mobile ions, either H$^+$/OH$^-$ from dissociated H$_2$O, or added salt, e.g., LiClO$_4$~\cite{CarradNL17}. Gating occurs by electric field driven ion migration with gate charge forming a concentric ion layer within $\sim 1$~nm of the nanowire surface~\cite{KimAdvMat13}. The resulting strong gating has seen electrolyte gating become a well-known approach to improved performance in materials ranging from organic semiconductors to chalcogenides~\cite{DuJMS15}. Electrolyte gating also provides a simpler route to achieving concentric gating action for nanowire devices~\cite{LiangNL12,CarradNL14}. The key result of this work is a demonstration that electrolyte-gated nanowire transistors remain functional even in the limit where the doping density becomes sufficiently high that traditional gating approaches fail. This is useful because heavy doping provides a path to reduced contact and channel resistance. Additionally, PE gates are far simpler to produce than traditional metal-oxide wrap-gate structures~\cite{PfundAPL06, DharaAPL11, StormNL12, BurkeNL15} and utilise an intrinsically biocompatible material~\cite{KrskoLangmuir03}. Nanopatterned PE gates have been used in applications from enacting external ionic doping of quantum devices~\cite{LiangNL12,FahlvikAFM15} to ionic-to-electronic signal transduction~\cite{CarradNL17}. Here we extend this to include improved $p$-type NWFETs for room temperature nanowire complementary circuits.

\begin{figure*}
\includegraphics[width=14cm]{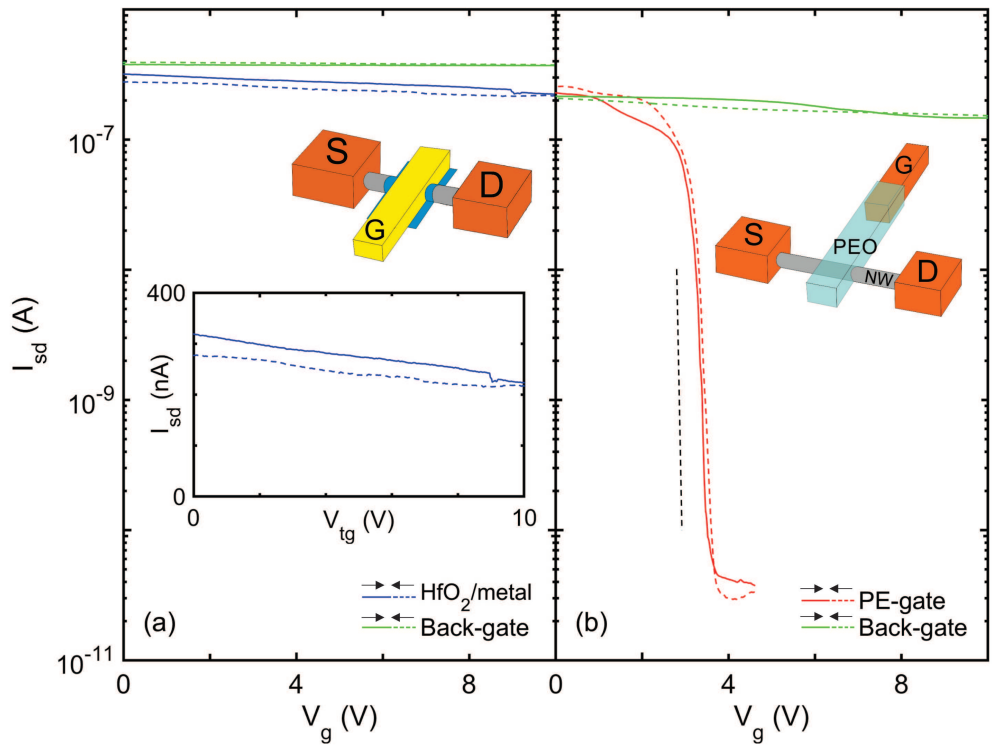}
\caption{Source-drain current $I_{sd}$ vs gate voltage $V_{g}$ for (a) metal-oxide top-gated and (b) PE-gated NWFETs. Both feature high Be acceptor concentration $N_A = 1.5 \times 10^{19}$~cm$^{-3}$ p$^+$-GaAs. $V_{sd} = 100$~mV for all traces with solid (dashed) lines showing sweeps towards more positive (negative) $V_{g}$. The blue and green traces in (a) are the metal top-gate and n$^+$-Si back-gates, respectively. The blue traces are offset downwards by $100$~nA for clarity. The green and red traces in (b) are for the n$^{+}$-Si back-gate and PE-gate respectively. The dashed black line in (b) indicates the thermal limit sub-threshold swing $60$~mV/dec. The inset to (a) shows the metal-oxide top-gated data from (a) on a linear scale to further demonstrate that weak gate modulation and no clear switching is observed. Corresponding device structures are inset to (a) and (b); S = source, D = drain, G = gate electrode, NW = nanowire, PEO = polyethylene oxide. The typical gate electrode G to nanowire gap is $2~\mu$m for the device in (b). All data obtained at room temperature.}
\end{figure*}

\section{Methods}
Our self-catalysed~\cite{ColomboPRB08} GaAs nanowires were grown by molecular beam epitaxy on (111)Si~\cite{CasadeiAPL13}. The undoped core was grown at $630^{\circ}$C using As$_4$ and a V/III flux ratio of $60$ for $30 - 45$ minutes. The Be-doped shell was grown at $465^{\circ}$C using As$_2$ and a V/III ratio of $150$ for $30$ minutes giving nanowires with typical diameter $150-200$~nm and length $5-7~\mu$m. The nanowires should be pure zincblende crystal phase throughout. There will possibly be some short wurtzite segments at the ends~\cite{KrogstrupNL10}; these will be buried under the contacts. We focus here on nanowires with shell acceptor density $N_{A} = 1.5 \times 10^{19}$~cm$^{-3}$; the highest doping density from our earlier work on all-inorganic $p$-GaAs NWFETs~\cite{UllahNanotech17}. Nanowires were transferred to a pre-patterned HfO$_2$/SiO$_2$-coated n$^+$-Si substrate for device fabrication with two architectures used: (a) a traditional metal-oxide $\Omega$-gate structure (Fig.~1(a) inset) and (b) a PE gate (Fig.~1(b) inset). Fabrication for both began with the contacts. These were defined by electron-beam lithography (EBL) and thermal evaporation of $200$~nm of $1\%$ Be in Au alloy (ACI Alloys). GaAs native oxide at the contact interfaces was removed by a $30$~s etch in $10\%$ HCl solution. The metal-oxide gate was produced in two steps. First a patterned $10$~nm layer of the high-$\kappa$ insulator HfO$_2$ was defined by EBL and atomic layer deposition (ALD). An overlapping $20$/$180$~nm Ti/Au gate electrode was then formed in a separate round of EBL and metal deposition.

\begin{figure*}
\includegraphics[width=14cm]{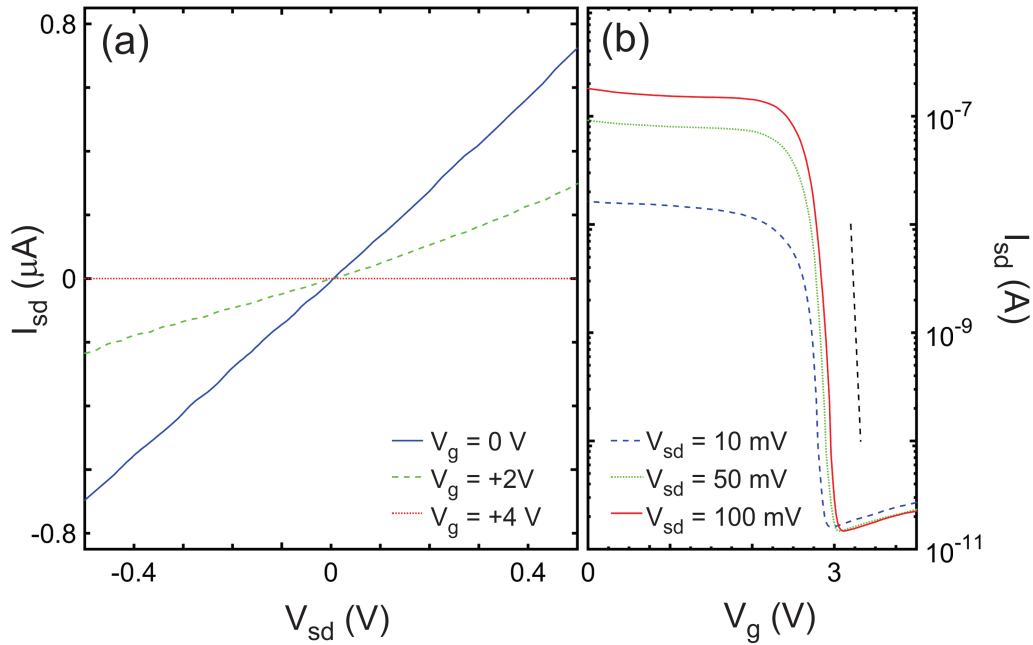}
\caption{(a) Source-drain current $I_{sd}$ vs source-drain voltage $V_{sd}$ for the PE-gated NWFET with $V_{g} = 0$~V (blue solid), $+2$~V (green dashed) and $+4$~V (red dotted) demonstrating the ohmic contact performance obtained via high acceptor density. (b) $I_{sd}$ vs PE gate voltage $V_{g}$ for $V_{sd} = 10$~mV (blue dashed), $50$~mV (green dotted) and $100$~mV (red solid) demonstrating slight dependence of threshold voltage on $V_{sd}$. Corresponding threshold voltage values are $2.75$~V, $2.80$~V and $2.90$~V, respectively. The dashed black line in (b) indicates the thermal limit sub-threshold swing $60$~mV/dec.}
\end{figure*}

For PE-gated devices, the Ti/Au gate electrode terminated $2~\mu$m from the nanowire and was defined by EBL and thermal evaporation~\cite{CarradNL14}. Patterning of the PE was the final step. $200$~mg polyethylene oxide (molecular weight $200$k -- Aldrich) was dissolved in $10$~mL methanol by sonication for $30$~min. Addition of LiClO$_4$ is optional and provides no significant performance enhancement in our NWFETs~\cite{CarradNL17}. For all of the PE gates used in this study, the polyethylene oxide is doped with LiClO$_4$ with a $1:10$ LiClO$_4$:PEO ratio. The solution was left to stand overnight with the supernatant spin-coated onto the device at $3000$~rpm for $60$~s. The device was baked at $90^{\circ}$C for 30~mins to remove residual methanol. Polyethylene oxide acts as a positive EBL resist~\cite{CarradNL14} and was directly patterned by EBL with beam energy $7$~keV and dose $300~\mu$C/cm$^2$. Development in H$_2$O removed unexposed regions giving nanoscale PE strips between the gate electrode and nanowire~\cite{CarradNL14}. Electrical measurements were performed at room temperature in ambient. Yokogawa GS200 voltage sources supplied the source-drain voltage $V_{sd}$ and gate voltage $V_g$ for both PE and metal-oxide gates. we measured the drain current $I_{d}$ using a Keithley 6517A electrometer.

\section{Results}
A key challenge for $p$-GaAs NWFETs is obtaining low contact resistance and strong gate performance simultaneously. Low resistance contacts to GaAs nanowires are difficult because surface-states pin the surface Fermi energy mid-gap, unlike for InAs, where the surface Fermi energy is pinned at the conduction band edge. This leads to a significant Schottky barrier for metal-GaAs interfaces. Ohmic contacts to GaAs typically use an annealed alloy of a noble metal and a diffusive dopant~\cite{BacaTSF97}, e.g., AuGe for $n$-type~\cite{MorkotterNL15} and AuBe for $p$-type~\cite{UllahNanotech17}. The idea is that the diffusive dopant causes the semiconductor local to the metal contact to become very highly doped. This makes the Schottky barrier depletion region very narrow, raising the electron tunnelling probability and giving a linear $I$-$V$ characteristic for the contact. This can be assisted by doping the semiconductor local to the contact by other means, e.g., by ion-implantation~\cite{StichtenothAPL08}, or during growth~\cite{GutscheJAP09,CasadeiAPL13}.

Our GaAs nanowires are readily doped $p$-type with Be, which incorporates preferentially via the nanowire side facets enabling structures with an undoped core and Be-doped shell~\cite{CasadeiAPL13}. A Be-doped shell provides an ideal interface for AuBe contacts with narrow Schottky barrier and thereby $p$-type NWFETs with low resistance ohmic contacts. We recently showed that contact annealing is detrimental in this case likely because the Be diffusion rate within the nanowire exceeds the Be out-diffusion rate from the alloy, reducing the net doping level at the contact interface~\cite{UllahNanotech17}. This provides a strong incentive to compensate by maximising shell doping density to ensure the Schottky barrier depletion width remains minimised. However, this approach brings a second problem: severe loss in gate performance. Figure~1(a) shows $I_{sd}$ versus $V_{g}$ for a $p$-GaAs nanowire MOSFET with high shell acceptor density $N_A = 1.5 \times 10^{19}$~cm$^{-3}$. The metal-oxide $\Omega$-gate (blue trace) modulates the current by a factor of $2$ at best despite the use of high-$\kappa$ dielectric. The data is presented on a focussed linear $I_{sd}$ scale in the inset to Fig.~1(a) to conclusively demonstrate this. To prove that this is not due to limited gate voltage range, we take a separate device to catastrophic gate dielectric breakdown without switching being observed. The data for this is shown in Supplementary Fig.~1. We obtain similar lack of gate efficacy from the n$^+$-Si back-gate (green trace). Poor gating at the high free carrier density arising from degenerate doping is expected; it is exactly why FETs do not have metal channels. An obvious potential solution is to keep the shell doping density high near the contacts yet lower near the gate. Unfortunately, it is not clear how to achieve nanowire shell growth with controlled axial doping variation. Ion implantation is an alternative but one that causes significant damage to the nanowire~\cite{StichtenothAPL08}. Thus for our $p$-GaAs nanowire MOSFETs we are trapped in an unfortunate trade-off between contact resistance and gate performance governed by shell doping density. Our key finding in this paper is that electrolyte gating offers a path to obtaining low contact resistance and good gate performance simultaneously.

Figure~1(b) shows the performance of the PE gate (red trace) for a $p$-GaAs nanowire with $N_A = 1.5 \times 10^{19}$~cm$^{-3}$ from the same growth. The green trace shows the performance for the $n^+$-Si back-gate to confirm that conventional gating still fails and there is no unanticipated difference between the nanowires in Figs.~1(a) and 1(b). The PE gate gives very strong gating with sub-threshold swing $S\sim75 \pm 15$~mV/dec and on-off ratio near $10^4$. The sub-threshold swing is comparable to the best obtained for $n$-InP ($68$~mV/dec)~\cite{StormNL11} and $n$-GaAs ($70$~mV/dec)~\cite{MorkotterNL15} nanowire MOSFETs and within $25\%$ of the room temperature thermal limit ($60$~mV/dec). We obtain on-current $I_{on}\sim0.25~\mu$A at $V_{sd} = 100$~mV corresponding to $400$~k$\Omega$ channel resistance, with contact resistance $R_{on} \sim 30$~k$\Omega$ previously measured for this doping level~\cite{UllahNanotech17}. Similar performance is obtained from separate nominally-identical devices as demonstrated by the data in Fig.~2. Figure~2(a) shows $I_{sd}$ versus $V_{sd}$ for a PE gated device at several $V_{g}$ demonstrating good linear contact performance throughout the entire PE gate range. Saturation is not observed for $V_{sd} < 2.5$~V in these high $N_{A}$ nanowires, as expected given their high doping density. A significant aspect of Fig.~1(b) is the low hysteresis, particularly in the sub-threshold regime, which we attribute to three factors. First, growth on (111)Si gives $\{110\}$ side-facets for which surface-states largely reside outside the band-gap~\cite{LinJVSTB12,ChelikowskyPRB79}. Second, there is no added oxide beyond the thin GaAs native oxide that grows upon air exposure. Third, the high density of free carriers in the shell and mobile ions in the PE strongly screen the residual surface state/oxide trapping effects. As with the $p$-GaAs nanowire MOSFETs, the precise performance is tunable via $N_A$, for example, at lower $N_A = 1\times10^{18}$~cm$^{-3}$ we see slightly poorer sub-threshold swing ($95$~mV/dec) and higher $R_{on} \sim 1.4$~M$\Omega$ but improved on-off ratio $\sim 10^6$, lower threshold voltage ($\sim+2$V) and no appreciable worsening in hysteresis. We also find that the threshold voltage is slightly influenced by $V_{sd}$ as shown in Fig.~2(b). We do not expect significant short-channel effects in our devices, and accordingly, the threshold shift direction is opposite that expected for drain-induced barrier lowering (DIBL). The threshold shift direction is instead indicative of the increased $I_{sd}$ that naturally follows increased $V_{sd}$ at fixed $V_{g}$, consistent with other nanowire transistors~\cite{StormNL11, DeyNL12}. Finally, we comment briefly on the field-effect mobility for our device. We can obtain the field-effect mobility $\mu_{FE} = g_{m}L^{2}/CV_{ds}$ with transconductance $g_{m} = \partial I_{ds}/\partial V_{g}$, channel length $L$ and gate capacitance $C$. The latter is difficult to accurately estimate for an electrolyte gate and should be considered an order-of-magnitude estimate at best. Here we estimate the capacitance of the electrical double layer at the PE/NW interface using a concentric cylinder formula $C = \frac{2\pi\epsilon_{r}\epsilon_{0}L}{\ln(1+t/r)}$ where $\epsilon_{0}$ is the permittivity of free space, $t$ is the electrical double layer thickness and $r$ is the nanowire radius. We measured $r~=~72.5$~nm by scanning electron microscopy, and $L~=~1.5~\mu$m, which is the length of nanowire covered by the PE gate, by optical microscopy. We take typical values of $\epsilon_{r} = 20$ and $t = 1$~nm for a polyethylene oxide formulation similar to ours from Takeya {\it et al.}~\cite{TakeyaAPL06} and thereby obtain $C \approx 0.12$~pF. This gives a capacitance per area of $17~\mu$F/cm$^{2}$, which is consistent with the $10~\mu$F/cm$^{2}$ typical of polyethylene oxide-based polymer electrolytes~\cite{KimAdvMat13}. With a measured transconductance $g_{m} = 175$~nS at $V_{sd} = 100$~mV for the data in Fig.~1(b), we obtain $\mu_{FE} \approx 0.3$~cm$^{2}$/Vs. The field-effect mobility is low compared to, e.g., InAs nanowire transistors, but this is not unexpected given the much higher effective mass $m^{*} \sim 0.35m_{0}$ for 1D-confined holes in GaAs~\cite{DanneauAPL06} (c.f., $m^{*} \sim 0.023m_{0}$ for electrons in InAs nanowires~\cite{ThelanderSSC04}), and the fact that conduction occurs largely via a thin, heavily-doped shell in our devices.

\section{Discussion}
We now put our results into context with other $p$-type III-V NWFETs. The most promising alternate III-V is GaSb, which is intrinsically $p$-type even when undoped due to native antisite defects~\cite{LingAPL04,VirkkalaPRB12}. Dey {\it et al.}~\cite{DeyNL12} recently reported on single InAs/GaSb nanowire CMOS inverters with a GaSb $p$-MOSFET sub-threshold swing $S = 400$~mV/dec, on-off ratio $\sim10^{1.8}$ and on-resistance $R_{on}~>~1.2$~M$\Omega$. Our device in Fig.~1(b) surpasses all three performance metrics. The on-resistance for GaSb NWFETs can be improved by Zn doping~\cite{BabadiAPL17} but this compromises on sub-threshold swing, as for GaAs $p$-MOSFETs~\cite{UllahNanotech17}. Babadi {\it et al.}~\cite{BabadiAPL17} obtain $R_{on}\sim26$~k$\Omega$ with $S\sim820$~mV/dec for moderate Zn doping and short channel length $L = 200$~nm but lose pinch-off for longer channels and/or higher doping levels. The one aspect where our PE-gated GaAs NWFETs fall behind is ac response. The InAs/GaSb CMOS inverter of Dey {\it et al.} shows square-wave fidelity loss at $\sim10$~kHz~\cite{DeyNL12}. We currently experience fidelity loss at $\sim10$~Hz due to the limited ionic conductivity of our PE, but our estimates suggest $\sim1$~kHz is possible with some engineering of the PE and device design~\cite{CarradNL17}, whilst MHz operation of other PE-gated devices is well established~\cite{KimAdvMat13}. A key limitation of polyethylene oxide-based electrolyte gates is their strong affinity for water and hygroscopic nature, which makes their performance sensitive to ambient humidity and hydration accumulated during processing~\cite{CarradNL17}. We expect improved performance to be obtained by a shift to other electrolyte-gate materials see, e.g., discussion in Kim {\it et al.}~\cite{KimAdvMat13}, as well as through further engineering of $N_A$ and the device architecture; this will be the subject of future work.

Briefly considering other $p$-type III-Vs, $p$-InAs gives ambipolar behaviour because conduction via the sub-surface electron layer from surface Fermi-level pinning competes with hole conduction in the core~\cite{SorensonAPL08}. This competition leads to poor off-current and sub-threshold slope performance albeit with low contact resistance at room temperature~\cite{SorensonAPL08}. $p$-InP is likewise ambipolar. The higher band-gap of InP aids with off-current suppression giving on-off ratio $>10^2$ with $S\sim220$~mV/dec but low $I_{on}~<~10$~pA~\cite{StormNL11}. In contrast, InSb has the smallest bandgap and gives the poorest performance in the role of room temperature $p$-NWFET~\cite{NilssonAPL10}.

\section{Conclusion}
We have shown that electrolyte gating remains effective in nanowire transistors where the doping level is sufficiently high that traditional metal-oxide gate formulations fail completely. We exploit this to obtain p-GaAs nanowire transistors that surpass other III-V p-type nanowire MOSFETs on a combination of three performance metrics: sub-threshold swing, on-off ratio and on-resistance. The latter is achieved via the high doping density which necessitates electrolyte gating for functional gating. We obtain a sub-threshold swing of $75 \pm 15$~mV/dec, within $25\%$ of the room-temperature thermal limit, and comparable with the best $n$-type nanowire MOSFETs. Additionally, our gate structures show low hysteresis in the sub-threshold regime, are easier to fabricate than metal-oxide gate structures, and feature an inherently biocompatible material. Our results point an interesting path to extending the performance and application of nanowire transistors, and motivate further work on improved electrolyte materials for nanoscale device and bioelectronics applications.

\begin{acknowledgments}
This work was funded by the Australian Research Council (ARC) Grants No. DP170104024 and No. DP170102552, the University of New South Wales, Danish National Research Foundation and the Innovation Fund Denmark. This work was performed in part using the NSW node of the Australian National Fabrication Facility (ANFF).
\end{acknowledgments}

\end{document}